\documentclass[prd,aps,floats,amssymb,nofootinbib,twocolumn]{revtex4}
\usepackage{epsfig}
\newcommand{\be}{\begin{equation}}
\newcommand{\ee}{\end{equation}}
\newcommand{\bea}{\begin{eqnarray}}
\newcommand{\eea}{\end{eqnarray}}
\newcommand{\lag}{{\cal L}}
\newcommand{\mpl}{M_{\rm Pl}}
\newcommand{\mn}{{\mu\nu}}

\begin{document}
\setlength{\unitlength}{1mm}
 
\title{Preheating in Derivatively-Coupled Inflation Models}

\author{Cristian Armendariz-Picon\footnote{armen@physics.syr.edu}, Mark Trodden\footnote{trodden@physics.syr.edu}, Eric J. West\footnote{ejwest@physics.syr.edu}}
\affiliation{Cosmology Group, Department of Physics, Syracuse University, Syracuse, NY 13244-1130, USA}

\begin{abstract}
We study preheating in theories where the inflaton couples derivatively to scalar and gauge fields. Such couplings may dominate in natural models of inflation, in which the flatness of the inflaton potential is related to an approximate shift symmetry of the inflaton. We compare our results with previously studied models with non-derivative couplings. For sufficiently heavy scalar matter, parametric resonance is ineffective in reheating the universe, because the couplings of the inflaton to matter are very weak. If scalar matter fields are light, derivative couplings lead to a mild long-wavelength instability that drives matter fields to non-zero expectation values.  In this case however, long-wavelength fluctuations of the light scalar are produced during inflation, leading to a host of cosmological problems. In contrast, axion-like couplings of the inflaton to a gauge field do not lead to production of long-wavelength fluctuations during inflation.  However, again because of the weakness of the couplings to the inflaton,  parametric resonance is not effective in producing gauge field quanta.
 
\end{abstract}
\maketitle

%==================================================================================================

\section{Introduction}

During a sufficiently long epoch of inflation, matter is diluted away by the quasi-exponential expansion of the universe.  The re-population of the universe with radiation after the end of inflation is then typically achieved by the decay of the inflaton into ordinary matter particles.  The perturbative theory of this process---{\it reheating}~\cite{Albrecht:1982mp,Abbott:1982hn}---was refined by the discovery~\cite{Dolgov:1989us,Traschen:1990sw,Kofman:1994rk} that in many models the dynamics would proceed through a stage of parametric resonance, leading to the extremely efficient decay of the inflaton into a far from equilibrium distribution of matter.  The understanding of this preliminary stage of reheating---{\it preheating}---has since been developed by many authors \cite{Shtanov:1994ce, Kofman:1995fi, Boyanovsky:1994me, Boyanovsky:1995em, Boyanovsky:1996sq, Greene:1997fu, Kaiser:1997mp, Khlebnikov:1996mc, Khlebnikov:1996wr, Kofman:1997yn}. 

Most work on preheating has focused on models with direct, non-derivative, couplings to matter.  Little is known about the strength of these couplings, but theoretical arguments suggest that they have to be rather weak. Indeed, in order for the inflaton to drive a phenomenologically acceptable stage of inflation, its potential has to be extremely ``flat". A variety of different ways to stabilize these flat potentials have been studied~\cite{Arkani-Hamed:2003mz}, but perhaps the most compelling idea that has emerged to date is that inflation is driven by a pseudo-Nambu-Goldstone boson~\cite{Freese:1990rb}. In the simplest realization of this idea, the inflaton sector is invariant under an approximate global $U(1)$ symmetry that shifts the inflaton field by a constant.  Because only a constant potential is invariant under shifts of the inflaton,  this approximate symmetry guarantees that deviations from flatness are small. But at the same time, because direct couplings between the inflaton and matter violate the shift symmetry, they are are also expected to be extremely weak, and perhaps even negligible.

On the other hand, derivative couplings of the inflaton to matter do satisfy the shift symmetry, and so there is no reason for them to be particularly weak. It is therefore entirely possible that derivative couplings could be more important than non-derivative ones during reheating.  Studies of preheating in models with derivative couplings~\cite{Dolgov:1994zq, Dolgov:1996qq} have been less extensive than those of direct couplings, and therefore our aim here is to carry out one such study: examining preheating in these models and determining whether they lead to a qualitatively different picture of the end of inflation and the onset of reheating. 

In the next section we will review the standard ideas of preheating in models with direct couplings between the inflaton and matter. In section~\ref{Derivative} we then motivate the study of derivatively coupled models and carry out the related preheating calculations, including couplings to both scalar and gauge fields, before concluding.

%==================================================================================================

\section{Parametric Resonance in Canonical Models}
\label{Canonical}

We begin by reviewing the basic results about preheating in a simple model with a direct coupling between the inflaton and matter fields.  For detailed methods and results regarding the physics of reheating and preheating, we refer the reader to~\cite{Kofman:1997yn,Bassett:2005xm}.

For simplicity we assume inflation is driven by a single real scalar field $\phi$, slow-rolling down a quadratic effective potential, and coupled to a massive real scalar field $\chi$ (representing matter fields) through a quartic interaction,
\be
	\lag = -\frac{1}{2}(\partial^\mu\phi)\partial_\mu\phi 
	       - \frac{1}{2}(\partial^\mu\chi)\partial_\mu\chi 
	       - \frac{m_\phi^2}{2}\phi^2 - \frac{m_\chi^2}{2}\chi^2
	       - \frac{g^2}{2}\phi^2\chi^2 \ .\label{eq:L}
\ee
Throughout we treat the inflaton field as spatially homogeneous and work in a spatially flat Friedmann-Robertson-Walker universe with metric ${ds^2=-dt^2+a^2(t)\, d\vec{x}^2}$.   

We expand the field $\chi$ in terms of annihilation and creation operators
\be
	\chi({\bf x},t) = \int\frac{d^3k}{(2\pi)^{3/2}}\left(a_k\,\chi_k(t)\,e^{i{\bf k}\cdot{\bf x}}
	+ \mathrm{h.c.}\, \right).
\ee
Then the mode functions $\chi_k$ and the homogeneous inflaton $\phi$ satisfy the field equations
\bea
  	\ddot\chi_k + 3H\dot\chi_k 
  	&+& \left[\left(\frac{k}{a}\right)^2 +m_\chi^2+ g^2\phi^2\right]\chi_k = 0,\label{eq:chi}
	\\
	\ddot\phi + 3H\dot\phi 
	&+& \left(m_\phi^2 +g^2 \chi_k^2\right)\phi = 0,\label{eq:phi}
\eea
where the Hubble parameter $H(t)\equiv{\dot a}/a$ is determined by the Friedmann equation  
\be
  	H^2 = \frac{1}{6\mpl^2}\left(\dot\phi^2 + m_\phi^2\phi^2\right),\label{eq:H}
\ee
and where we have denoted a time derivative by an overdot. Note that we have neglected the effects of the $\chi$ field and its interactions with $\phi$ on the expansion rate of the universe, since we are interested in an epoch at which the inflaton is dominant, so that $\rho_\chi\ll\rho_\phi$. Similarly, we will henceforth also neglect the back-reaction term $g^2\chi^2\phi$ in equation~(\ref{eq:phi}).

Slow-roll inflation occurs when ${\dot\phi \ll m_\phi \phi}$ and ${\ddot\phi \ll 3H \dot\phi}$. During slow-roll,  the interaction between $\phi$ and $\chi$ in equation~(\ref{eq:chi}) simply increases the effective mass of the field $\chi$,
\be \label{eq:effective mass}
	m_\mathrm{eff}^2= m_\chi^2+g^2\phi^2.
\ee
There are several arguments that suggest that this mass should not be much lower than the Hubble scale. First, if $m_\mathrm{eff}\ll H$, a non-zero value of $\chi$ remains frozen during inflation, so there is no reason to assume that $\chi$ lies at the minimum of its potential. At the same time, if the effective mass of $\chi$ is smaller than the Hubble scale, long-wavelength fluctuations of $\chi$ created during the inflationary epoch generically lead to isocurvature perturbations. And finally,  if $m_\mathrm{eff}\ll H$, the same long-wavelength fluctuations will drive $\chi$ away from zero, even if $\chi=0$ initially, which could eventually lead to a second stage of inflation driven by the field $\chi$ \cite{Felder:1999pv}.  If $m_\chi$ is negligible, the requirement $H<m_\mathrm{eff}$, in combination with equation (\ref{eq:H}), leads to the bound
\be\label{eq:g bound}
	g\gtrsim\frac{m_\phi}{\mpl} \approx 10^{-6}.\label{eq:coupling bound}
\ee
The same bound also guarantees that the effective mass of $\chi$, equation (\ref{eq:effective mass}), is adiabatically constant during inflation, even if $m_\chi$ is negligible. Hence, the coupling between the inflaton and any other light scalar cannot be too weak. 

Once inflation ends, the inflaton field begins oscillating around the minimum of its effective potential. During this oscillation stage, the expansion of the universe mimics a matter domination phase. In particular, soon after the end of inflation, the field evolution follows the equation
\be
	\phi(t)\approx \phi_\mathrm{e}\, \frac{\sin (m_\phi t)}{m_\phi t},
\ee
where $\phi_\mathrm{e}\approx \mpl$ denotes the value of the inflaton at the beginning of the oscillating stage\footnote{More precisely, the value of the inflaton at the beginning of the oscillating phase is actually smaller, $\phi_\mathrm{e}\approx 0.3\mpl$.}, at $t\approx m_\phi^{-1}.$  Because of the rapid oscillations, the effective mass of $\chi$ ceases being adiabatically constant for sufficiently light fields.  One may then ask whether this non-adiabatic evolution leads to particle production.  The answer is affirmative for appropriate couplings between matter and the inflaton.  More importantly,  parametric resonance can excite $\chi$ very efficiently. 

If we neglect the expansion of the universe and the back-reaction from $\chi$, equations (\ref{eq:phi}) and (\ref{eq:chi}) can be combined into the Mathieu equation,
\be \label{eq:Meq}
	\ddot\chi_k + \left[\delta - 2\epsilon\cos(2t)\right]\chi_k=0,\label{eq:Mathieu}
\ee
where we have defined a dimensionless time $t\to m_\phi^{-1}t$, and used
\be
	\delta \equiv \left(\frac{k}{m_\phi}\right)^2 + \left(\frac{m_\chi}{m_\phi}\right)^2 
               + 2\epsilon \quad \text{and} \quad
	\epsilon \equiv \left(\frac{g \, \phi_\mathrm{e}}{2m_\phi}\right)^2.
\ee

Solutions of the Mathieu equation are known to exhibit parametric resonance---resonance for certain values of the dimensionless parameters $\delta$ and $\epsilon$.  In the $\delta$-$\epsilon$ plane, these resonant solutions form band-like patterns called instability bands (see~\cite{Nayfeh:1981}). Along these unstable solutions, the mode functions grow exponentially,
\be
	\chi_k\propto\exp(\mu \, m_\phi \, t ), 
\ee
where the characteristic exponent $\mu$ depends on $\delta$ and $\epsilon$.

\begin{figure}[h]
  \includegraphics[scale=0.7]{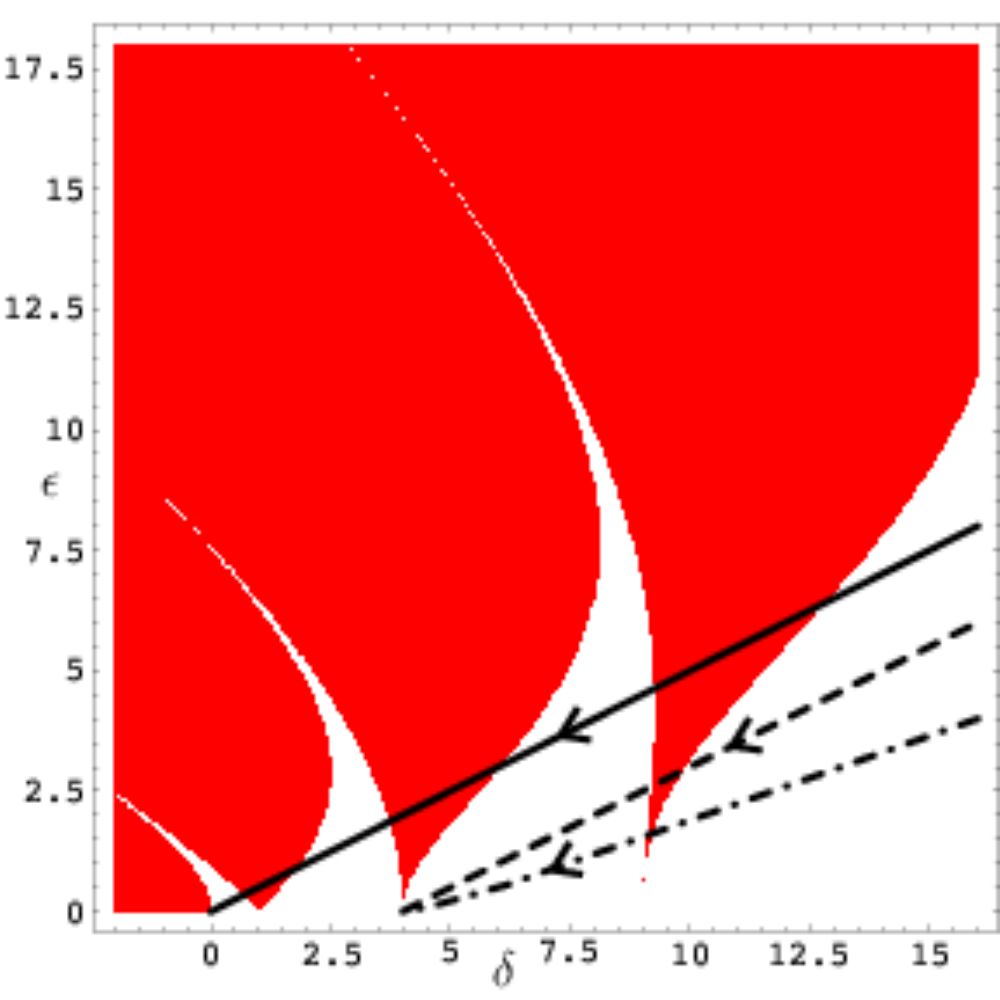}
  \caption{Instability diagram for solutions of the Mathieu equation in the $\delta$-$\epsilon$ plane.  In this diagram, shaded regions indicate resonant, or unstable, solutions of the Mathieu equation.  Unshaded regions indicate stable solutions. We also show the evolution of the parameters in equation~(\ref{eq:parameters1}) for different comoving modes and fields: $k=0, m_\chi=0$ (continuous), $k=0, m_\chi\neq 0$ (dashed) and $k\neq 0, m_\chi\neq 0$ (dot-dashed). }
  \label{Meq_fig1}
\end{figure}

\begin{figure}[h]
  \includegraphics[scale=0.4]{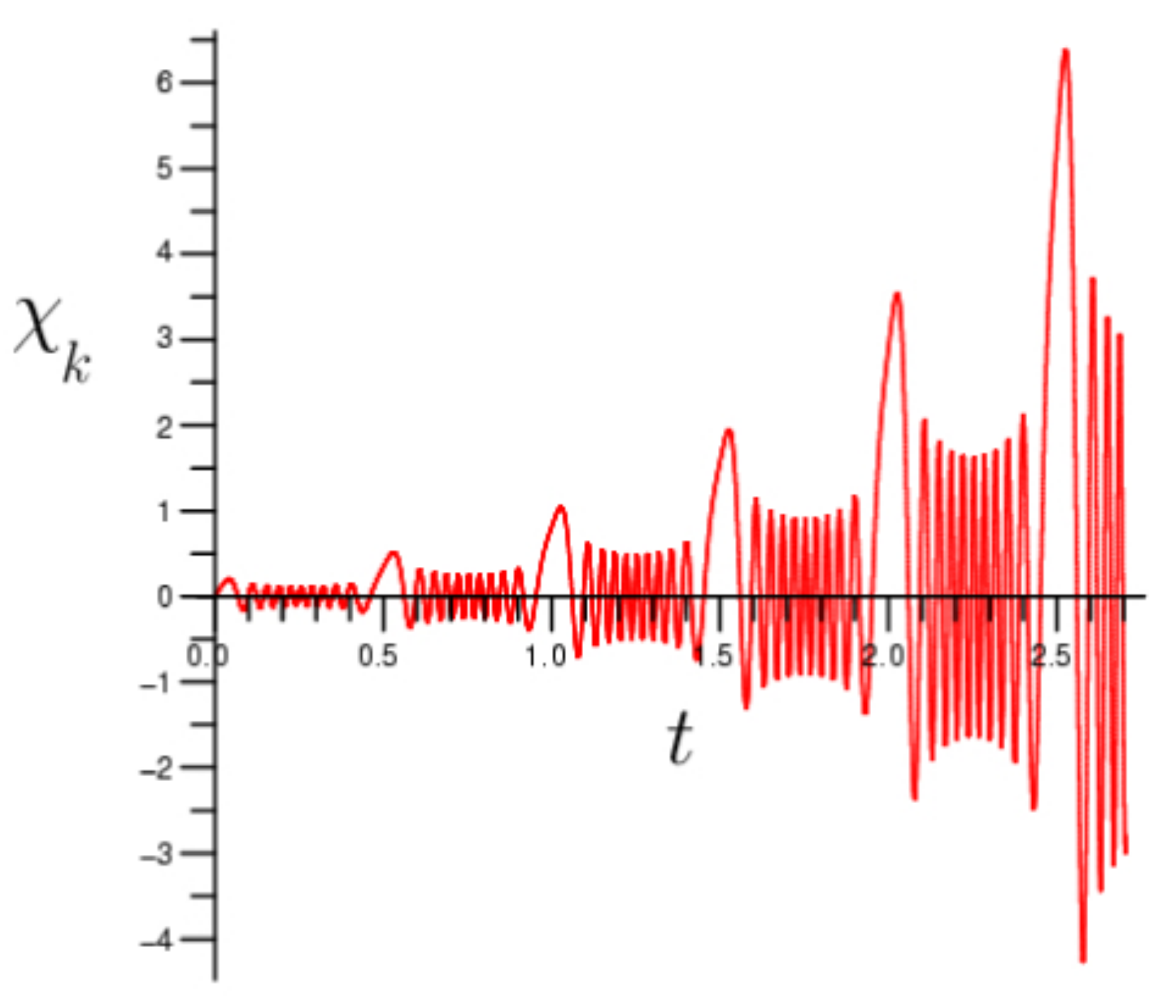}
  \includegraphics[scale=0.4]{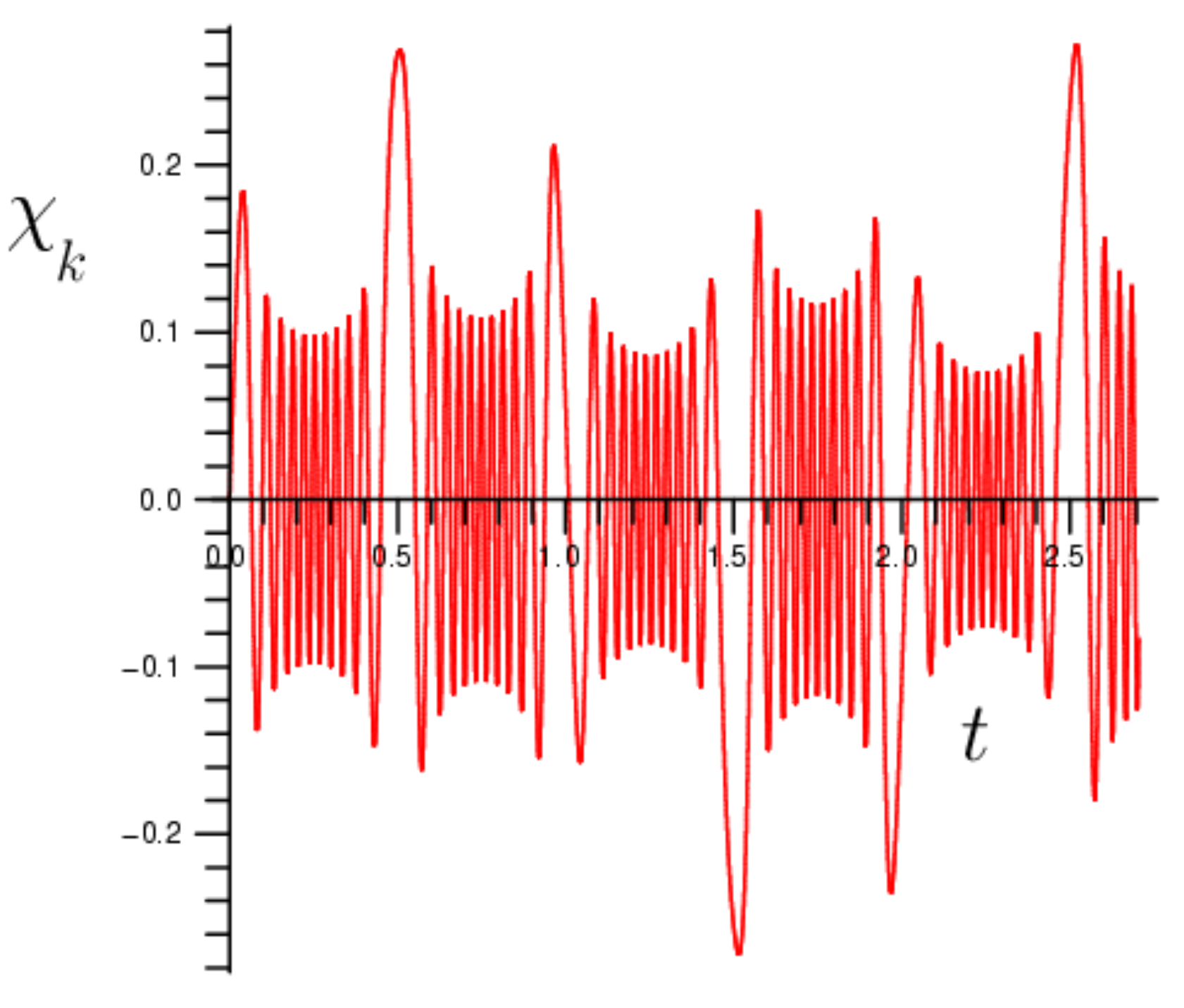}
  \caption{Top panel: Time evolution of an unstable mode when expansion of the universe is neglected. Here $m_\chi=0$, $k=m_\phi$, and $g^2=1\times 10^{-3}$.\\
  Bottom panel: Time evolution of a stable mode when expansion of the universe is neglected. Here $m_\chi=0$, $k=3 m_\phi$, and $g^2=1\times 10^{-3}$. Here, and in all remaining figures,  time is measured in units of $2\pi/m_\phi$ which corresponds to the number of oscillations in $\phi$ from the end of inflation.}
  \label{C1_fig}
\end{figure}

Although the Mathieu equation cannot be solved exactly for all values of $\delta$ and $\epsilon$, using perturbative techniques in the regime of  small $\epsilon$, one can show that the tips of the instability bands occur at $\delta=n^2$ where $n$ is an integer~\cite{Landau:1976,Nayfeh:1981}.  A numerical plot of the bands is shown in~Fig.~\ref{Meq_fig1}.  Extrapolating the tips of the plotted instability bands down to the $\epsilon=0$ axis shows agreement with the above perturbative results. In the theory we consider here however, equation (\ref{eq:g bound}) implies $\epsilon\gtrsim 1$. Therefore, preheating occurs in the  broad-resonanace regime, which, as we shall see, is quite different from its narrow-resonance counterpart at $\epsilon\lesssim 1$. In~Fig.~\ref{C1_fig} we give examples of the behavior of unstable and stable modes.  Similar results and a more in-depth discussion can be found in~\cite{Kofman:1997yn}.

\begin{figure}[h]
  \begin{center}
  \includegraphics[scale=0.4]{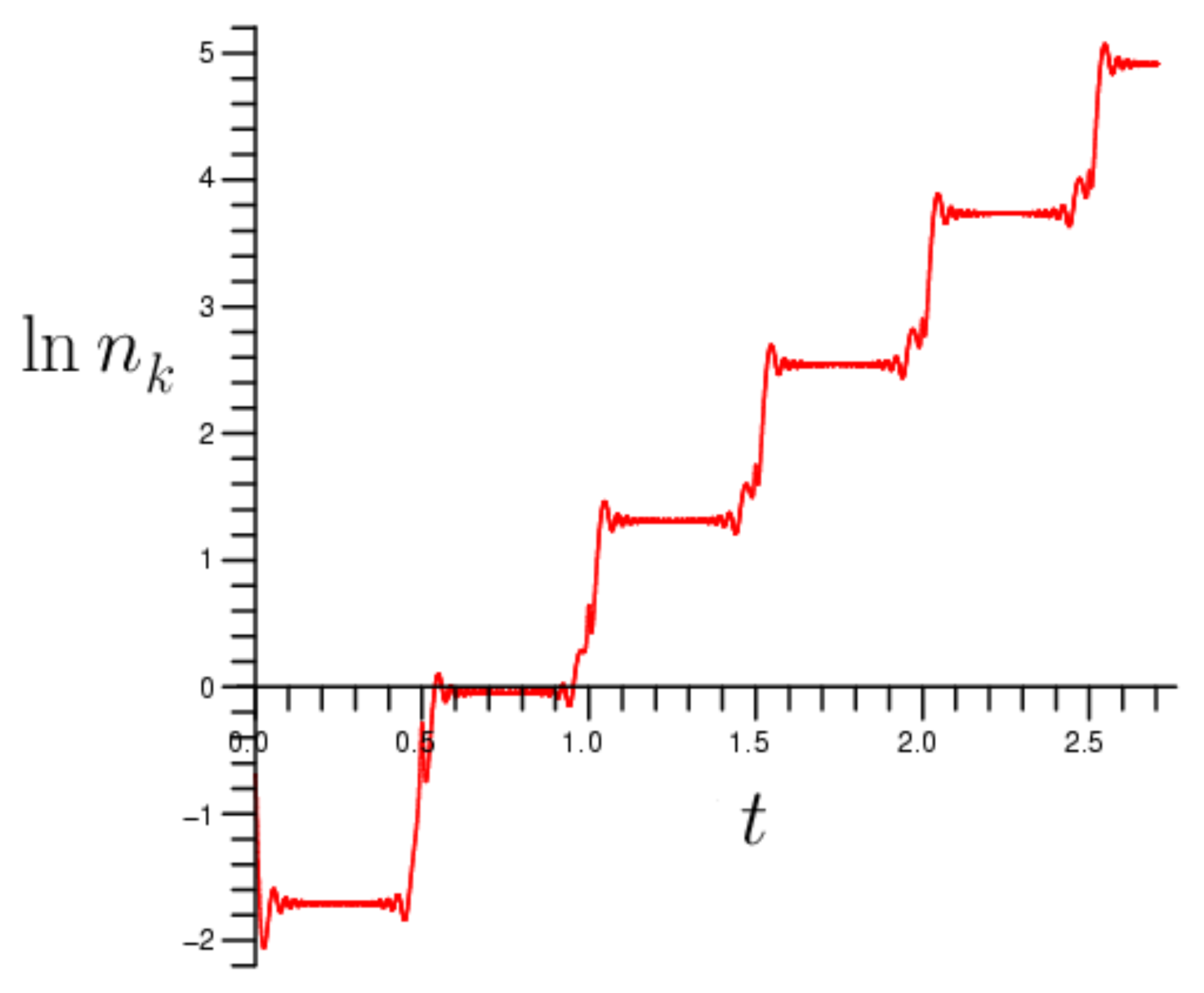}
  \end{center}
  \caption{Comoving number density of $\chi$-particles in an unstable mode when expansion of the universe is neglected. Here, as in the top panel of Fig.~\ref{C1_fig}, $m_\chi=0$, $k=m_\phi$, and $g^2=1\times 10^{-3}$.}
  \label{C1_ndensity1_fig}
\end{figure}

Since we have neglected the expansion of the universe, the parameters of the Mathieu equation  are time-independent.  A mode that lies within one of the instability  bands  therefore continues to grow indefinitely. Because the effective  particle number density $n_{k}$ is related to the mode functions by
\be
	n_k = \frac{1}{2\omega_k}\left(|\dot\chi_k|^2 + \omega_k^2|\chi_k|^2\right) 
	- \frac{1}{2},\label{eq:ndensity}
\ee
where $\omega_k^2=m_\mathrm{eff}^2+(k/a)^2$, parametric resonance leads to unbounded growth of particle number, as shown in Fig.~\ref{C1_ndensity1_fig}.  Of course this unending particle production is simply a consequence of neglecting any dissipative effects on the inflaton (the back-reaction of matter and the expansion of the universe).

When the expansion of the universe is included, the inflaton undergoes damped oscillations resulting in a loss of energy with which to drive parametric resonance in the matter fields.   But on the other hand,  the expansion of the universe  redshifts all momenta, which causes the energy of a mode to decrease as the universe expands. It is possible to see both of these effects analytically by making the approximation $a\propto t^{2/3}$ and substituting into equations~(\ref{eq:phi}) and~(\ref{eq:chi}). Then, the rescaled field 
\be\label{eq:rescaled}
	\tilde{\chi}_k\equiv \chi_k\,  a^{3/2}
\ee
satisfies the Mathieu equation (\ref{eq:Meq}), but now the parameters $\delta$ and $\epsilon$ become time-dependent
\be\label{eq:parameters1}
  	\delta = \left(\frac{k}{m_\phi a}\right)^2 + \left(\frac{m_\chi}{m_\phi}\right)^2 
	       + 2\epsilon \quad \text{and} \quad
  	\epsilon = \left(\frac{g\phi_\mathrm{e}}{2m_\phi t}\right)^2.
\ee
Even though $\delta$ and $\epsilon$ are time-dependent, their values are nearly constant during the oscillating phase. For instance, the relative change in $\epsilon$ during an oscillation of $\phi$ is
\be
	\frac{1}{m_\phi}\frac{\dot{\epsilon}}{\epsilon} = -\frac{3H}{m_\phi}.
\ee
Hence, at late times, $t\approx H^{-1}\gg m_\phi^{-1}$ we can think of the parameters $\delta$ and $\epsilon$ as being (locally) constant. The same assumption of late times has been actually made in the derivation of (\ref{eq:parameters1}).  

\begin{figure}[h]
  \begin{center}
  \includegraphics[scale=0.4]{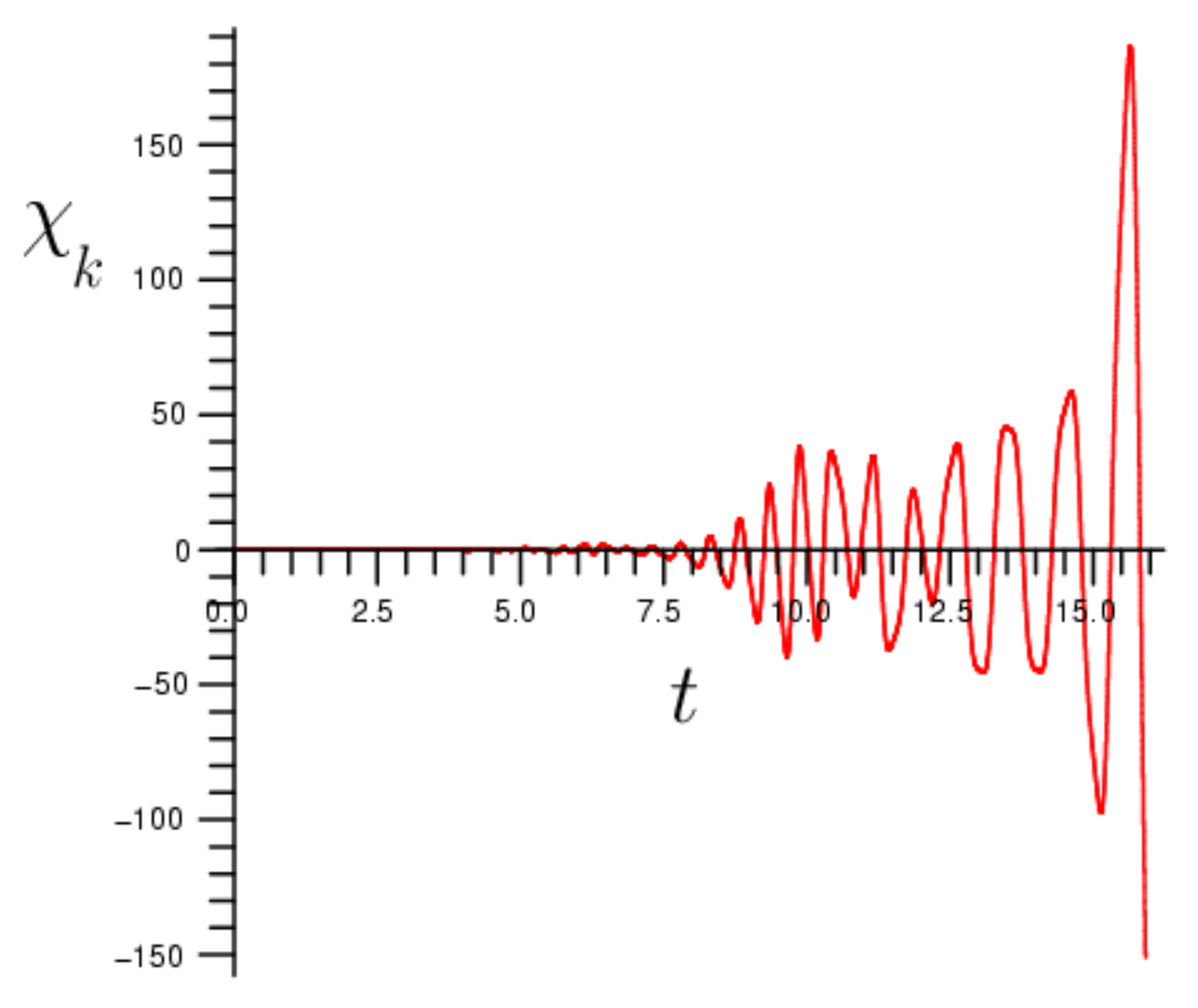}
  \includegraphics[scale=0.4]{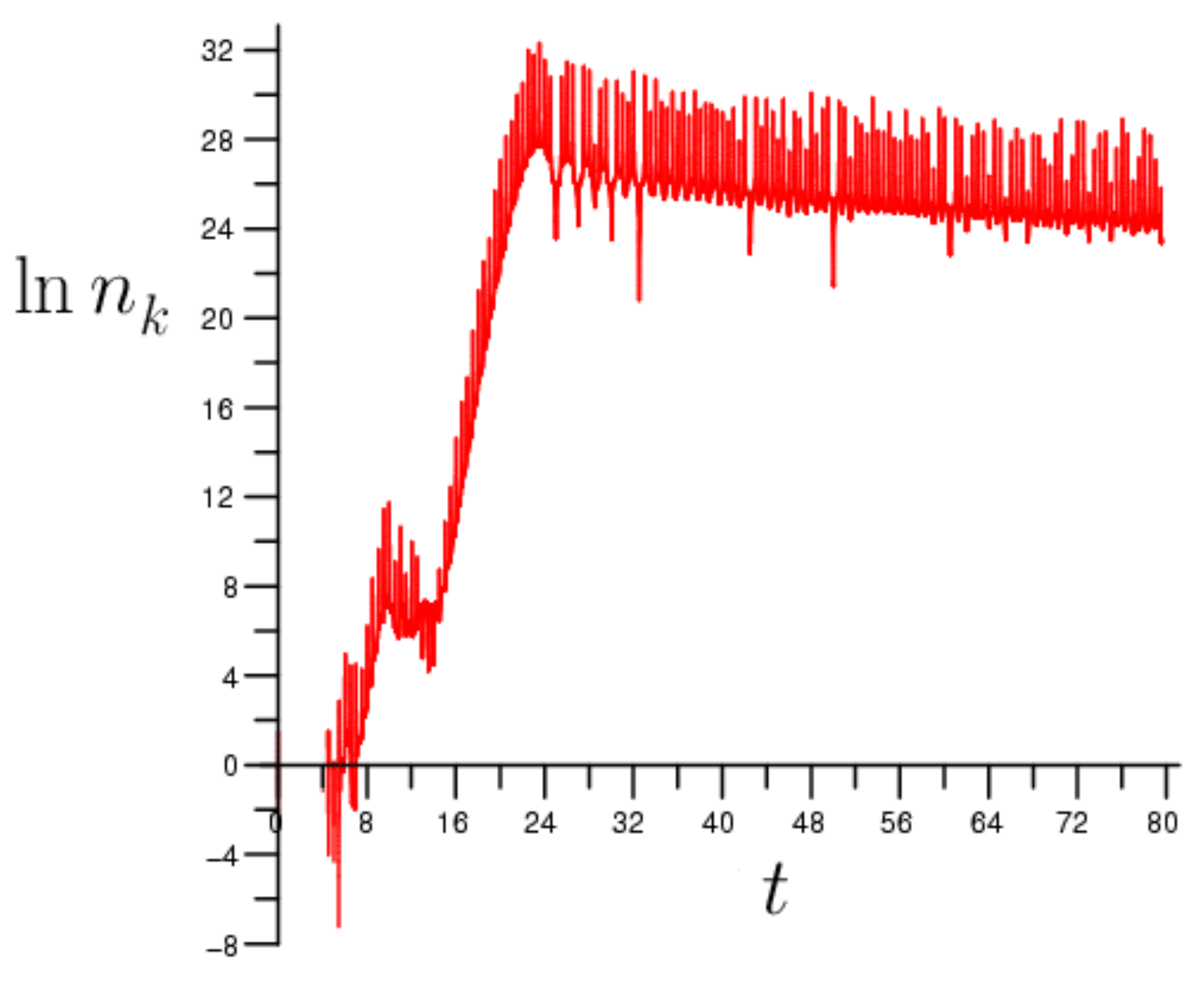}
  \end{center}
  \caption{Top panel: Time evolution of an unstable mode when expansion of the universe is included. Here $m_\chi=0$, ${k=10^{-1}\times m_\phi}$, and $g^2=1\times 10^{-2}$.  The resonance is less explosive than when expansion is neglected, and it only lasts briefly as the mode passes through an instability band.  After the mode passes through the band, it begins to decay due to the expanding background. \\
Bottom panel: Comoving number density of $\chi$-particles in the same mode as top panel.  Here it is clearly seen that the mode passes through two instability bands (the first between $t\simeq 8$ and $t\simeq 12$, the second between $t\simeq 15$ and $t\simeq 30$).  While it passes through each band the comoving number density increases exponentially.}
  \label{C1_unstable2a_fig}
\end{figure}

The time-dependence in $\delta$ and $\epsilon$ will cause comoving modes to migrate in the $\delta$-$\epsilon$ plane.  Thus, when the expansion of the universe is taken into account, particle production in a given $k$-mode may take place for a brief time interval as this mode passes in and then out of an instability band.  We also plot the trajectories of different modes in the $\delta$-$\epsilon$ plane in Fig. \ref{Meq_fig1}. The longer a particular mode remains inside an instability band, the more efficient is the production of the corresponding particles. It is clear from the figure that parametric resonance is always more effective for longer wavelengths and lighter fields \cite{Zlatev:1997vd}. If the initial value of $\epsilon$ is small, the trajectory of, say, the $k=0$ mode only crosses the narrow tip of the first instability band; this is the narrow-resonance regime, which is not particularly effective in producing particles.  On the other hand, if the initial value of $\epsilon$ is sufficiently high, comoving modes remain inside  several instability bands a significant fraction of the time. We illustrate the behavior of unstable modes in an expanding background in Fig.~\ref{C1_unstable2a_fig}. Note that parametric resonance can excite modes of $\chi$ with energies much higher than the inflaton mass if $\epsilon$ is sufficiently big.

Other models with canonical couplings between the inflaton and matter can be similarly examined.  For example, one can include trilinear terms proportional to $\phi\chi^2$ that would arise naturally in theories in which $\phi$ has a non-vanishing vacuum expectation value.  Although the quantitative results differ, similar qualitative results emerge.  What is less explored is whether models with non-canonical couplings between the inflaton and matter yield similar results.  In the next section we examine such couplings.

%==================================================================================================

\section{Reheating in Models with Derivative Couplings}
\label{Derivative}

\subsection{Derivative Couplings in Inflationary Scenarios}

Phenomenologically viable inflationary models require very ``flat" potentials. A simple example is provided by our chaotic model with quadratic potential ${V(\phi)=\frac{1}{2}m^2\phi^2}$, in which sufficient flatness implies that the mass of the field must be $m\approx 10^{-6}\mpl$. If the inflaton did not couple to matter, this value would not pose any particular problem, as no quantum corrections (other than gravitational) could drive $m$ to higher values. However, the inflaton must couple to matter for reheating to occur, and these couplings will generically be responsible for large quantum corrections.  In the absence of any symmetry, a light inflaton hence requires fine tuning. 

\begin{figure}[h]
  \includegraphics{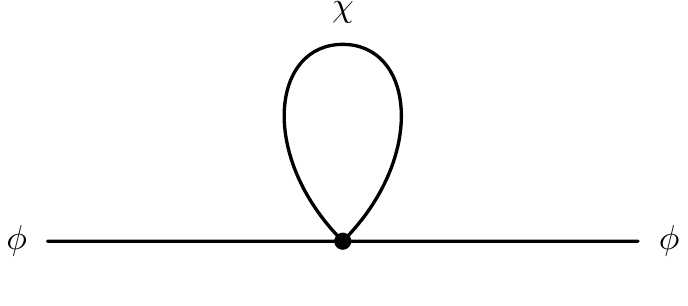}
  \caption{One-loop quantum correction to the mass of the inflaton. A quartic coupling between the inflaton $\phi$ and matter $\chi$ leads to a quadratically divergent mass.}
  \label{fig:loop}
\end{figure}

To be more specific consider an inflationary  model with  bare Lagrangian given by equation (\ref{eq:L}). To lowest order in $g^2$ the effective mass of $\phi$ receives a correction given by the Feynman diagram in Fig. \ref{fig:loop}. The loop integral is quadratically divergent, so the corrected mass squared is
\be
	m^2\approx m_\phi^2+\frac{g^2 \Lambda^2}{16\pi^2},
\ee
where $\Lambda$ is the ultraviolet cut-off of the theory, which is expected to be Planckian.  There are hence two different ways to obtain ${m\approx 10^{-6}\mpl}$: (i) The coupling of the inflaton to matter is strong, $g\approx 1$,  and there has to be a cancellation between the bare mass and the quantum corrections of 1 part in $10^6$, or (ii) the inflaton is very light and it couples very weakly to matter. Case  (i) requires fine tuning; as we shall see, case (ii) is realized in models where the inflaton is a Pseudo-Goldstone boson. Incidentally, let us point out that if quantum corrections to the mass of the inflaton are to be small, $g^2\Lambda^2/8\pi^2<m_\phi^2$, the bound in equation (\ref{eq:g bound}) has to be violated.

Because of the fine tuning required in case (i), significant efforts have been devoted to develop inflationary models in which the inflaton is naturally light~\cite{Arkani-Hamed:2003mz}. As in many standard model extensions, these models require additional symmetries.  Perhaps the most promising candidates are models in which the inflaton is a pseudo-Nambu-Goldstone boson, the Goldstone boson of an approximate, spontaneously broken symmetry~\cite{Freese:1990rb}. In the limit that the symmetry is exact, the Goldstone boson is massless, and the inflaton potential is exactly flat.  Small symmetry breaking terms then give a correspondingly small mass to the inflaton, and the remaining approximate symmetry guarantees that quantum corrections remain small.\footnote{It is sometimes argued that quantum gravity effects, such as the creation of virtual black holes, explicitly break this symmetry \cite{Kallosh:1995hi}. We shall ignore this possibility here.}

As we have discussed above, if the inflaton is to decay into matter fields, it is, of course, important that it couple to matter. In the context of natural models, this generically implies that matter fields must be charged under the spontaneously broken symmetry. To begin our exploration of reheating in natural models of inflation, let us therefore consider the simplest model that satisfies both requirements: a theory with two charged complex fields $\Phi$ and $\chi$, invariant under a spontaneously broken (approximate) $U(1)$ symmetry, 
\bea
	\lag = &-& (\partial_\mu\Phi) \partial^\mu\Phi^* 
	        - (\partial_\mu \chi) \partial^\mu \chi^*-(F^2-\Phi^* \Phi)^2 \nonumber
\\
     	       &-& m_\chi^2 \chi^* \chi 
                - \lambda(\Phi^2 \chi^2+\Phi^*{}^2 \chi^*{}^2).\label{eq:L2}
\eea
This Lagrangian is invariant under the $U(1)$ transformation $\Phi\to e^{i\phi}\Phi$, $\chi\to e^{-i\phi}\chi$, where $\phi$ is a complex number. This phase will later play the role of the inflaton, while $\chi$ will play the role of matter. For simplicity, we also assume that the theory has an additional $\mathbb{Z}_2$ symmetry $\chi\to -\chi$.

The $U(1)$ symmetry is spontaneously broken, because in the vacuum the field $\Phi$ has a non-vanishing expectation value, $\langle|\Phi|\rangle=F$. As we shall argue below, in order for the inflaton potential to be sufficiently flat, we need $F\approx\mpl$. To study the low-energy excitations around the vacuum, we redefine
\be
  	\Phi\to F\exp\left(i\frac{\phi}{F}\right), \quad 
  	\chi\to \chi\exp\left(-i\frac{\phi}{F}\right),\label{eq:redefinition}
\ee
and ignore ``radial" excitations around the minimum of the potential because they are extremely heavy---with masses of order $F$. Substituting these expressions into~(\ref{eq:L2}) we find 
\bea
  	\lag = &-& (\partial_\mu\phi) \partial^\mu\phi - (\partial_\mu\chi) \partial^\mu\chi^*
       	        - m_\chi^2\chi^*\chi \nonumber 
\\
      	       &-& \lambda F^2(\chi^2 + \chi^*{}^2) 
      	        - \frac{1}{F^2}(\partial_\mu\phi) (\partial^\mu\phi)|\chi^2| \nonumber
\\
    	       &-&\frac{i}{F}(\partial_\mu\phi)
    	       \left(\chi^*\partial^\mu\chi - \chi\partial^\mu\chi^*\right).\label{eq:L3}
\eea
Note that this automatically generates derivative couplings between the inflaton $\phi$ and the field $\chi$ (which could be redefined away only if $\lambda=0$). These couplings are model-independent, in the sense that they originate from the kinetic term of the complex field $\chi$ and the field redefinition~(\ref{eq:redefinition}).

For $\phi$ to be a viable inflaton, it is necessary to generate a potential for it. To this end, we introduce terms in the Lagrangian  that explicitly break the $U(1)$ symmetry.  Because the potential has to vanish in the limit of exact symmetry, these terms will generically lead to an inflaton potential of the form
\be
  	V(\phi) \approx \mu^4 \left[1-\cos\left(\frac{\phi}{F}\right)\right],\label{eq:potential}
\ee
where $\mu$ is a parameter with dimensions of mass that characterizes the strength of the symmetry breaking terms. Note that the potential has to be periodic, since $\phi/F$ is a phase. For $\phi/F<1$, we can assume that the potential is quadratic, with squared mass $m_\phi^2=\mu^4/F^2$, which is the form we shall consider in the following.   

It is convenient to work with real fields, rather than complex ones. Defining
\be
  	\chi = \chi_r + i\chi_i,
\ee
where $\chi_r$ and $\chi_i$ are, respectively, the real and imaginary parts of $\chi$, the Lagrangian~(\ref{eq:L3}) becomes
\bea
	\lag = &-& (\partial_\mu\phi) \partial^\mu\phi - (\partial_\mu\chi_r) \partial^\mu\chi_r
    	        - (\partial_\mu\chi_i) \partial^\mu\chi_i
\\
     	       &-& (m_\chi^2 + 2\lambda F^2)\chi_r^2 
     	        - (m_\chi^2 - 2\lambda F^2)\chi_i^2 \nonumber
\\
	       &+& \frac{2}{F}(\partial_\mu\phi)
      	       \left(\chi_r\partial^\mu\chi_i - \chi_i\partial^\mu\chi_r\right)
		- \frac{1}{F^2}(\partial_\mu\phi)(\partial^\mu\phi)\,(\chi_r^2 + \chi_i^2) \nonumber .
\eea
Therefore, derivative couplings between the inflaton and matter are quite generic and come in two classes: cubic and quartic. Note that in order to avoid a tachyonic instability, we must assume $m_\chi^2>2\lambda F$.   It is also easy to verify that the potential~(\ref{eq:potential}) will satisfy the slow-roll conditions only for $F\gtrsim \mpl$. 
It is unclear whether we can trust an effective field theory description with $F\gg\mpl$, so we shall implicitly assume $F\approx M_P$. Since $\mpl$ denotes the reduced Planck mass, we could push this choice a bit higher.   

\subsection{Parametric Resonance}
\label{Resonance}

In section~\ref{Canonical} we described how the inflaton may decay by resonantly exciting modes of its decay products. This process is non-perturbative in nature and requires us to solve the equations of motion of the matter fields $\chi$ during the oscillating stage of the inflaton at the end of inflation. We now wish to examine the extent to which this phenomenon may occur in models with derivative couplings, such as~(\ref{eq:L3}).

Because, by assumption, the fields $\chi$ have a vanishing expectation value at the end of inflation, the cubic terms in the interaction will have no effect on their equation of motion. Thus, we may concentrate on the quartic interaction,
\bea
  	\lag = -\frac{1}{2}(\partial^\mu\phi)\partial_\mu\phi 
  	      &-& \frac{1}{2}(\partial^\mu\chi)\partial_\mu\chi - \frac{1}{2}m_\phi^2\phi^2 
  	       -\frac{1}{2}m_\chi^2\chi^2 \nonumber 
\\
 	      &-& \frac{1}{F^2}(\partial^\mu\phi)(\partial_\mu\phi)\chi^2.\label{eq:L4}
\eea
The resulting equations of motion are
\bea
	\left[1 + 2\left(\frac{\chi}{F}\right)^2\right]\nabla_{\mu}\nabla^{\mu}\phi 
        &=& -4\frac{\chi}{F^2}(\partial_{\mu}\phi)\partial^{\mu}\chi + m_{\phi}^2\phi, \nonumber
\\
	\nabla_{\mu}\nabla^{\mu}\chi 
        &=& 2\frac{\chi}{F^2}(\partial_{\mu}\phi)\partial^{\mu}\phi + m_{\chi}^2\chi,
\eea
with the mode equation for $\chi$ therefore given by
\be
  	\ddot\chi_k + 3H\dot\chi_k 
	+ \left[\left(\frac{k}{a}\right)^2 + m_\chi^2 
	        - \frac{2}{F^2}\dot\phi^2\right]\chi_k = 0.\label{eq:chi4}
\ee

The form of this equation already hints that inflation and preheating in derivatively coupled models may be qualitatively different from the canonical case considered in section~\ref{Canonical}. This is because the field $\chi$ has an effective squared mass given by 
\be
  	m_\mathrm{eff}^2 = m_\chi^2-\frac{2}{F^2}\dot{\phi}^2.
\ee
Although $\dot{\phi}^2/F^2$ is small during slow-roll inflation,  the correction term becomes large when the inflaton leaves the slow-roll regime. Therefore, for sufficiently small values of $m_\chi$, the effective mass of the field $\chi$ might become tachyonic. Clearly,  such an instability is absent as long as the effective mass is positive, 
\be\label{eq:no tachyon}
	m_\chi \gtrsim \frac{\sqrt{2}m_\phi \mpl}{F}.
\ee
Interestingly, this bound is very similar to the requirement that the mass of $\chi$ be larger than the Hubble scale during inflation. As a matter of fact, since $F\approx \mpl$, the bound (\ref{eq:no tachyon}) simply reads $m_\chi\gtrsim m_\phi$. During slow-roll $H\approx m_\phi \, \phi/\mpl$, and because $\phi$ is of order $\mpl$ during the period of inflation accessible to present observations, equation (\ref{eq:no tachyon}) also follows from requiring $m_\chi>H$. It turns out that the tachyonic instability associated with the violation of the bound (\ref{eq:no tachyon}) is actually harmless, and it is just the production of long-wavelength fluctuations of the field $\chi$ what renders  light scalar fields undesirable. 

To further study parametric resonance in this class of models, it will be instructive to recast equation~(\ref{eq:chi4}) into the form of the Mathieu equation.  Including the expansion of the universe but neglecting back-reaction from $\chi$,  we find that the rescaled variable (\ref{eq:rescaled}) satisfies the Mathieu equation (\ref{eq:Mathieu}), with 
\be\label{eq:parameters2}
	\delta = \left(\frac{k}{m_\phi a}\right)^2 + \left(\frac{m_\chi}{m_\phi}\right)^2
	       - 2\epsilon \quad \text{and} \quad 
	\epsilon = \frac{1}{2}\left(\frac{\phi_\mathrm{e}}{F t}\right)^2.
\ee

\begin{figure}[t]
  \includegraphics[scale=0.7]{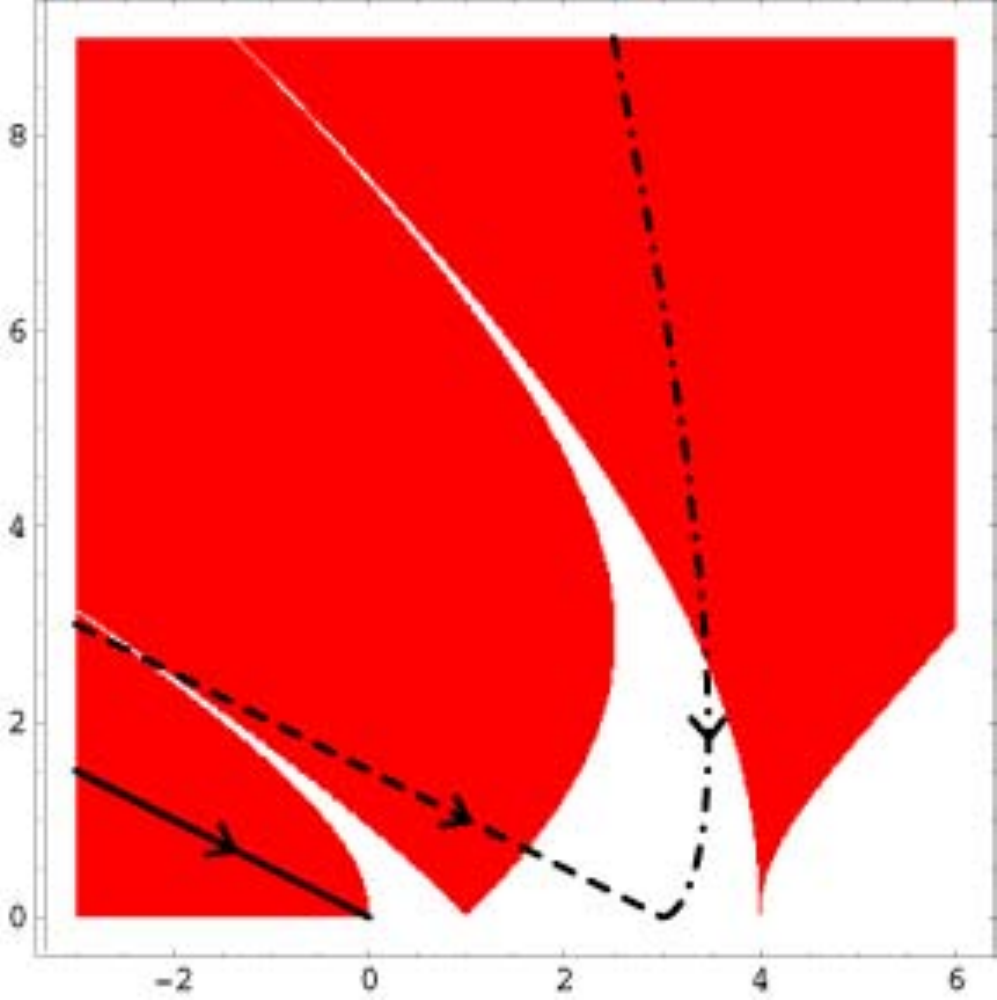}
  \caption{Instability diagram for solutions of the Mathieu equation in the $\delta$-$\epsilon$ plane.  We superimpose the evolution of the parameters in equation (\ref{eq:parameters2}) for different comoving modes and fields: $k=0, m_\chi=0$ (continuous), $k=0, m_\chi\neq 0$ (dashed) and $k\neq 0, m_\chi\neq 0$ (dot-dashed). }
  \label{Meq_fig2}
\end{figure}
The diagram in Fig.~\ref{Meq_fig2} shows different trajectories of a mode in the $\delta$-$\epsilon$ plane. It is again apparent from the figure that for a given initial value of $\epsilon$ preheating is in general more efficient for longer wavelength modes and lighter fields.  
A crucial property of derivatively coupled models is that $\epsilon$ is at most of order one, because $\phi_\mathrm{e}\approx \mpl\approx F$. Hence,  any resonance proceeds  close to the end of the instability bands, and is clearly ineffective if  fields are heavy,  ${m_\chi \gg m_\phi \, (\delta\gg 1)}$. On the other hand, if condition (\ref{eq:no tachyon}) is violated, long-wavelength modes will remain inside the instability band as long as the effective mass is negative, which could lead to a significant growth of the homogeneous component of $\chi$.   

\begin{figure}[t]
  \includegraphics[scale=0.4]{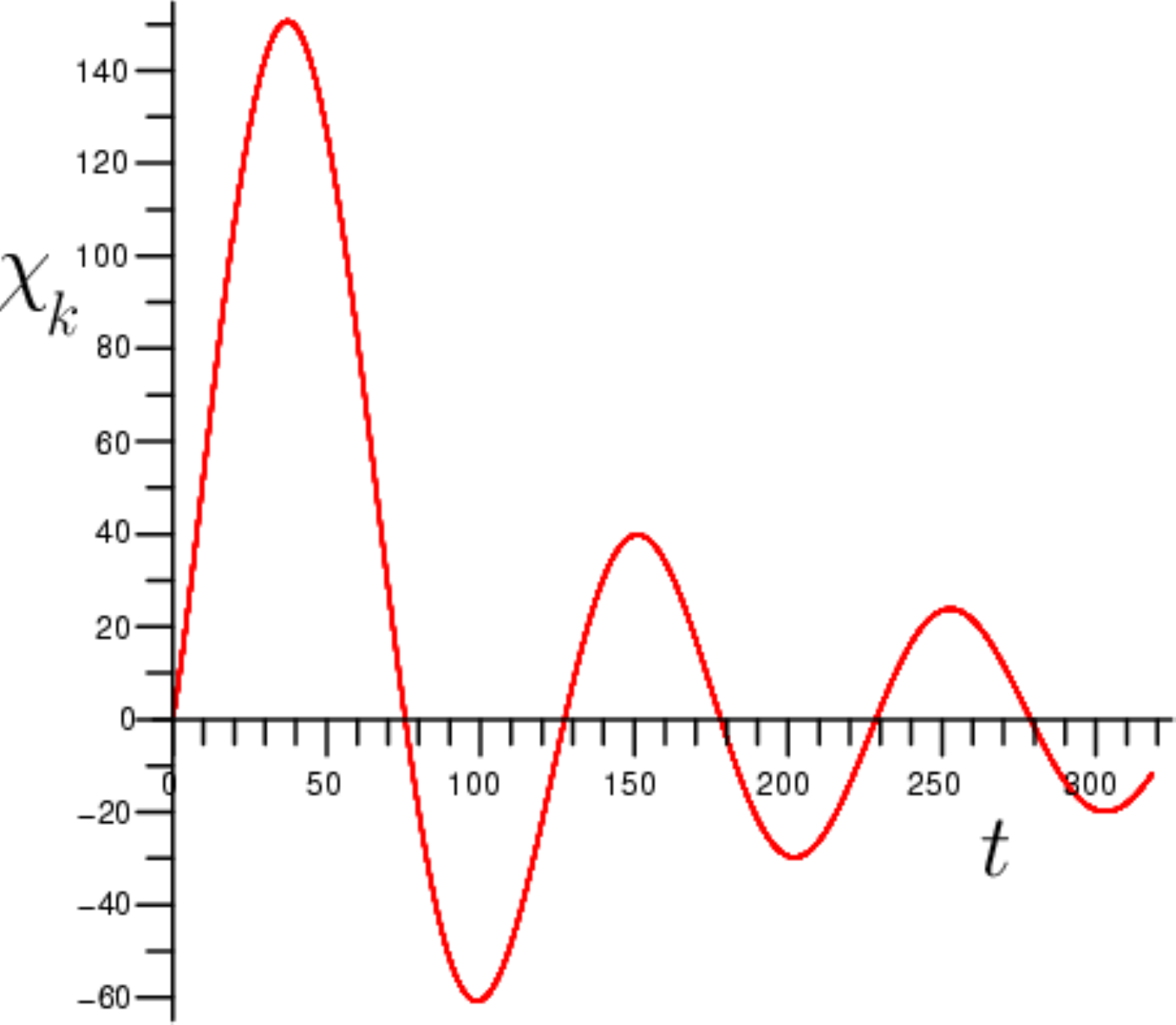}
  \caption{The zero mode of $\chi$  for $m_\chi=10^{-2}m_\phi$. During the initial stages, the effective mass is tachyonic, and the mode grows as it remains in the stability band at $\delta<0$. Once the mode exits the tachyonic regime, the mode function decays. Note that long-wavelength excitations do not admit a particle interpretation. }
  \label{fig:D2_stable}
\end{figure}

In order to estimate the efficiency of preheating in the tachyonic regime, let us consider the evolution of the ${k=0}$ mode of a  light field,  $m_\chi\ll m_\phi$. As we argued above, this is the case where parametric resonance is strongest. After the end  of inflation, the mode $k=0$ finds itself in the instability band at $\delta<0$. To lowest order in $\epsilon$, the characteristic exponent is simply $\mu=\sqrt{-\delta}$. Hence, for adiabatic changes in $\mu$, we expect the solutions to grow as 
$\exp (\int\mu\,   m_\phi\,  dt)$. However, this is only a good approximation if $\phi_\mathrm{e}/F\gg 1$. Indeed the adiabaticity parameter is
\be
	\frac{\dot{\mu}}{\mu^2} = -\frac{F}{\phi_\mathrm{e}},
\ee 
which is of order one instead. Therefore, we shall solve equation (\ref{eq:Meq}) directly by approximating
\be
	\delta=-\left(\frac{\phi_\mathrm{e}}{F t}\right)^2,\quad \epsilon=0.
\ee
The solution is a linear combination of powers in $t$. Keeping the growing mode, and returning to the original variable, we find
\be
	\chi\approx \chi_\mathrm{e} \left(\frac{t}{t_\mathrm{e}}\right)^p,
	\quad \text{where} \quad
	p=-\frac{1}{2}+\sqrt{\frac{1}{4}+\frac{\phi_\mathrm{e}^2}{F^2}},
\ee
and $\chi_\mathrm{e}$ denotes the value of $\chi$ at the end of inflation.
Clearly, as seen in Fig. (\ref{fig:D2_stable}), for  $\phi_\mathrm{e}/F\approx 1$ this amplification is very modest. Hence, parametric resonance  is rather ineffective for derivatively coupled scalar fields, no matter whether heavy or light. 

\subsection{Preheating into Gauge Fields}

Our analysis so far indicates that in natural models of inflation the population of  the universe with particles faces several challenges. First,  if matter fields are heavier than the inflaton, parametric resonance is absent, and the inflaton cannot decay perturbatively into matter for kinematical reasons. And second, if matter fields are lighter than the inflaton, parametric resonance or perturbative decay are possible, but then  long-wavelength fluctuations of matter are produced during inflation.   These conclusions crucially hinge on the assumption that matter fields can be represented by minimally coupled scalar fields, which, of course, is not the case. 
 
There are just two other types of matter fields the inflaton can couple to: fermions (spin 1/2 and 3/2) and vectors (spin 1). Reheating into fermions is not very effective because of the exclusion principle \cite{Greene:2000ew}, so we are left with vector fields as the only remaining alternative. Incidentally, vector fields are particularly appealing as decay products of the inflaton for two reasons: Their masses are protected by gauge symmetries, so they are naturally light, and the conformal nature of their couplings to gravity prevents the production of  long-wavelength fluctuations of these fields during inflation. 

Consider therefore an effective theory in which a light pseudo-scalar $\phi$ (the inflaton) couples derivatively to a $U(1)$ gauge field $A_\mu$ (matter), 
\be
	\lag = -\frac{1}{2}(\partial_\mu\phi)\partial^\mu\phi - \frac{1}{2}m_\phi^2\phi^2 
	       - \frac{1}{4}A^{\mn}A_{\mn} - \frac{\phi}{4F}\tilde{A}^{\mn}A_{\mn}.\label{eq:L5}
\ee
Here ${A_{\mu\nu}=\partial_\mu A_\nu-\partial_\nu A_\mu}$ is the field strength of the gauge fields, $\tilde{A}_{\mu\nu}\equiv \varepsilon^{\mu\nu\rho\sigma}A_{\rho\sigma}/2$ is its dual tensor, and $F$ is a symmetry breaking scale, with dimensions of mass. Despite the  appearances, the field $\phi$ couples derivatively to the gauge field, because $\tilde{A}_{\mu\nu}A^{\mu\nu}$ is a total derivative. Effective theories of this type appear in axion-like models \cite{Kim}, in which the coupling of the scalar to the electromagnetic field arises from the triangle anomaly. This sort of  coupling between a  pseudoscalar and  electromagnetism has been suggested as a mechanism to generate primordial magnetic fields in the early universe \cite{Turner:1987bw}, although a detailed study of coherent magnetic field production during inflation and the subsequent reheating stage has concluded that this generation is not sufficiently strong on the required scales \cite{Garretson:1992vt}. 

To analyze the dynamics of reheating in this model, we choose to work in the transverse gauge (where $\partial_i A^i=0$ and $A^0=0$).  In this gauge, the equations of motion for the inflaton and the non-vanishing components of $A_\mu$ are
\bea
	\nabla_\mu\nabla^\mu\phi
	&=&m_\phi^2\phi 
        + \frac{1}{a^4 F}\epsilon^{ijk}\dot{A}_i\partial_j A_k \label{eq:phi and A}
\\
	\eta^{\mu\nu}\delta^{ij}\partial_\mu A_{\nu j} 
	&=& \frac{1}{F}\epsilon^{ijk}
	[\dot\phi\,\partial_j A_k - (\partial_j\phi)\dot{A}_k].\label{eq:modefxn}
\eea
We expand $A_i$ in terms of mode functions:
\be
  	{\bf A}({\bf x},t) 
	= \int\frac{d^3k}{(2\pi)^{3/2}}
	\sum_{r=\pm}{\bf e}_r({\bf k}) A_r({\bf k},t)\, e^{i{\bf k}\cdot{\bf x}},
\ee
where ${\omega_k=|{\bf k}|\equiv k}$ is the  dispersion relation of the photons, and ${{\bf e}_r({\bf k})}$ are circular polarization vectors, ${{\bf k}\times {\bf e}_\pm= \mp\,  i\, k \, {\bf e}_\pm}$.  Substituting this expansion into (\ref{eq:modefxn}) yields the decoupled mode equations
\be
	\ddot{A}_\pm + H\dot{A}_\pm 
	+ \left[\left(\frac{k}{a}\right)^2 \mp \frac{k}{a}\frac{\dot\phi}{F}\right]A_\pm
	= 0.\label{eq:A}
\ee

\begin{figure}[h]
  \includegraphics[scale=0.7]{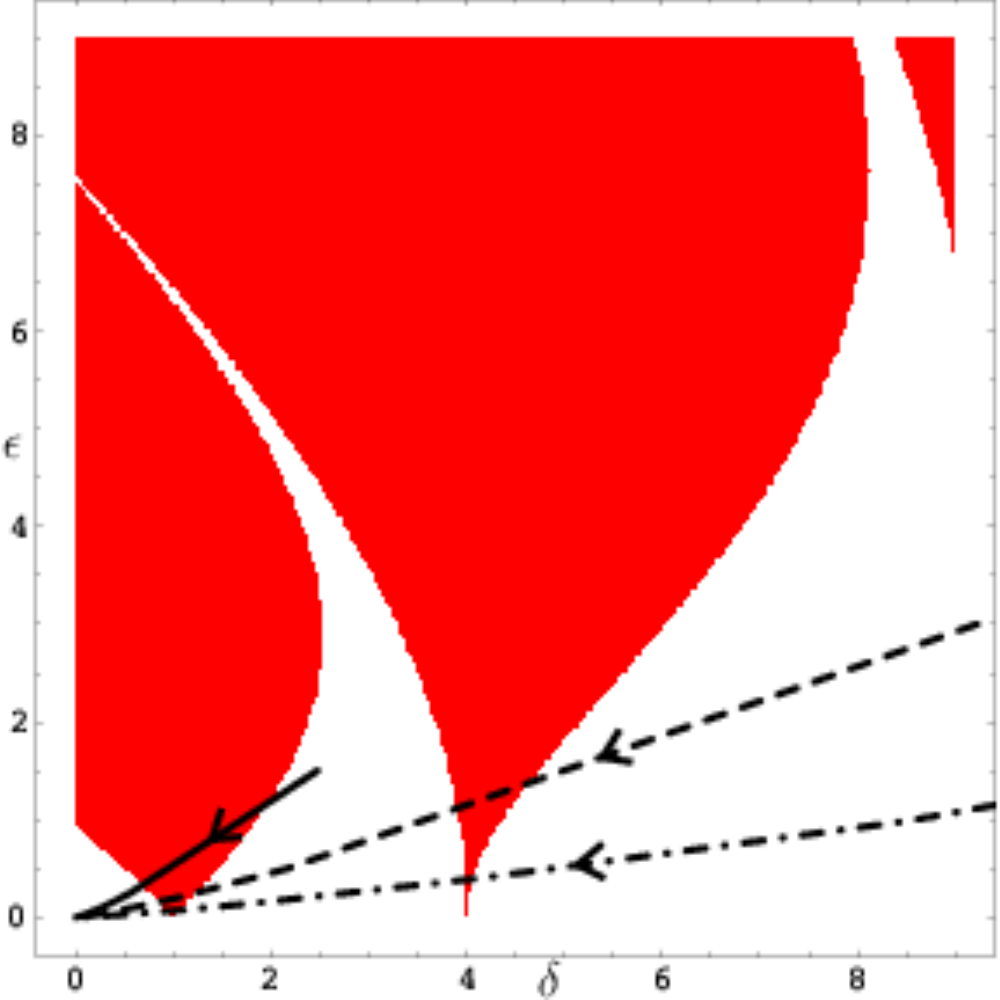}
  \caption{Instability diagram for solutions of the Mathieu equation in the $\delta$-$\epsilon$ plane.  The trajectories follow the evolution of the parameters in equation (\ref{eq:parameters3}) starting at the end of inflation. The corresponding values of $k_\mathrm{e}\equiv  k/(m_\phi a_\mathrm{e})$ are  $k_\mathrm{e}=3/4$ (continuous), $k_\mathrm{e}=3/2$ (dashed) and $k_\mathrm{e}=3$ (dot-dashed). In all cases $\phi_\mathrm{e}/F=1$.}
  \label{Meq_fig3}
\end{figure}

As was the case in the previously considered models, equation (\ref{eq:A}) exhibits parametric resonance for certain values of $k$ and $F^{-1}$, which can be interpreted as explosive particle production (e.g., photon production).  This can be seen analytically in the regime where back-reaction is neglected by putting (\ref{eq:A}) into the form of the Mathieu equation (\ref{eq:Mathieu}). The rescaled field variable,
\be
	\tilde{A}_{\pm}\equiv A_\pm \, a^{1/2}
\ee
 satisfies the Mathieu equation (\ref{eq:Meq}) with parameters
\be\label{eq:parameters3}
	\delta = \left(\frac{2k}{m_\phi a }\right)^2+\frac{2}{9t^2}, \quad \text{and} \quad 
	\epsilon= \pm\frac{k}{m_\phi a}\,  \frac{1}{t} \frac{\phi_\mathrm{e}}{F},
\ee  
where we have introduced the dimensionless time variable $t\to 2m_\phi^{-1}t.$ There is a significant difference that distinguishes couplings to gauge fields from those to scalar fields. Whereas for the latter the value of $\epsilon$ is  constant in a non-expanding universe, for the former it is proportional to the  energy of the photons. Hence, $\epsilon$ can become very large. Since the effective theory we are using breaks down at physical momentum $k/a \approx F$ we find that at the end of inflation $\epsilon$ can be as large as $\phi_\mathrm{e}/m_\phi\approx 10^{6}.$  However, this does not necessarily imply that parametric resonance is very effective. In Fig. \ref{Meq_fig3} we show the trajectories of different modes in the $\delta$-$\epsilon$ plane for different values of $k$ and $\phi_\mathrm{e}/F=1$. As seen in the figure, although a mode can cross several instability bands,  it does so in the narrow band regime, close to the tips of the instability bands, mainly because $\epsilon$ rapidly decays as time evolves. 

In order to estimate the efficiency of preheating in this regime, let us estimate how long a mode remains in the first instability band (the one whose tip touches $\epsilon=0,\,  \delta=1$), and how much its mode function grows during that time. For small $\epsilon$, the characteristic exponent along the first band is
\be\label{eq:coefficient}
	\mu=\frac{1}{2}\sqrt{\epsilon^2-(\delta-1)^2}.
\ee
Hence, its boundary is $|\delta-1|=\epsilon$, along which $\mu=0$. Let us write the time at which a mode enters and leaves the band as $t_1\mp \Delta t$, where $t_1$ is the time at which the mode crosses the center of the band, at $\delta=1$. Using equations (\ref{eq:parameters3}), and neglecting the factor $2/9 t^2$, we  find 
\be
	t_1\approx t_\mathrm{e} \left(\frac{2 k}{m_\phi a_\mathrm{e}}\right)^{3/2},
	\quad
	\Delta t\approx\frac{3}{8}\frac{\phi_\mathrm{e}}{F},	
\ee
where $t_\mathrm{e}\approx 1$ is the time at the end of inflation, and $a_\mathrm{e}$ denotes the value of the scale factor at that time.

We shall estimate now what is the growth of the mode function during that time. Inside the band, we can neglect the term $(\delta-1)^2$ in equation (\ref{eq:coefficient}), so that $\mu \approx \epsilon_1/2$, where $\epsilon_1$ is the value of $\epsilon$ at time $t_1$. Hence, we find that $\tilde{A}_+$ grows by a factor
\be\label{eq:growth}
	\exp (\epsilon_1\cdot \Delta t) \approx
	\exp\left[\frac{3}{16 t_\mathrm{e}}\frac{\phi_\mathrm{e}^2}{F^2}\left(\frac{m_\phi a_\mathrm{e}}{2 k}\right)^{3/2}\right],
\ee
while the growth in $A_+$ is suppressed by an additional factor $(t_\mathrm{e}/t)^{1/3}$.  Note that the above expression is strictly valid only for $\Delta t \ll t_1$, and provided the change in ${\mu=\epsilon/2}$ at time $t_1$ is adiabatic,
\be\label{eq:non-adiabatic}
	\frac{\dot{\mu}}{\mu^2}\bigg|_{t_1}\approx  -\frac{20}{3}\frac{F}{\phi_\mathrm{e}}\ll 1.
\ee

\begin{figure}[t]
  \includegraphics[scale=0.4]{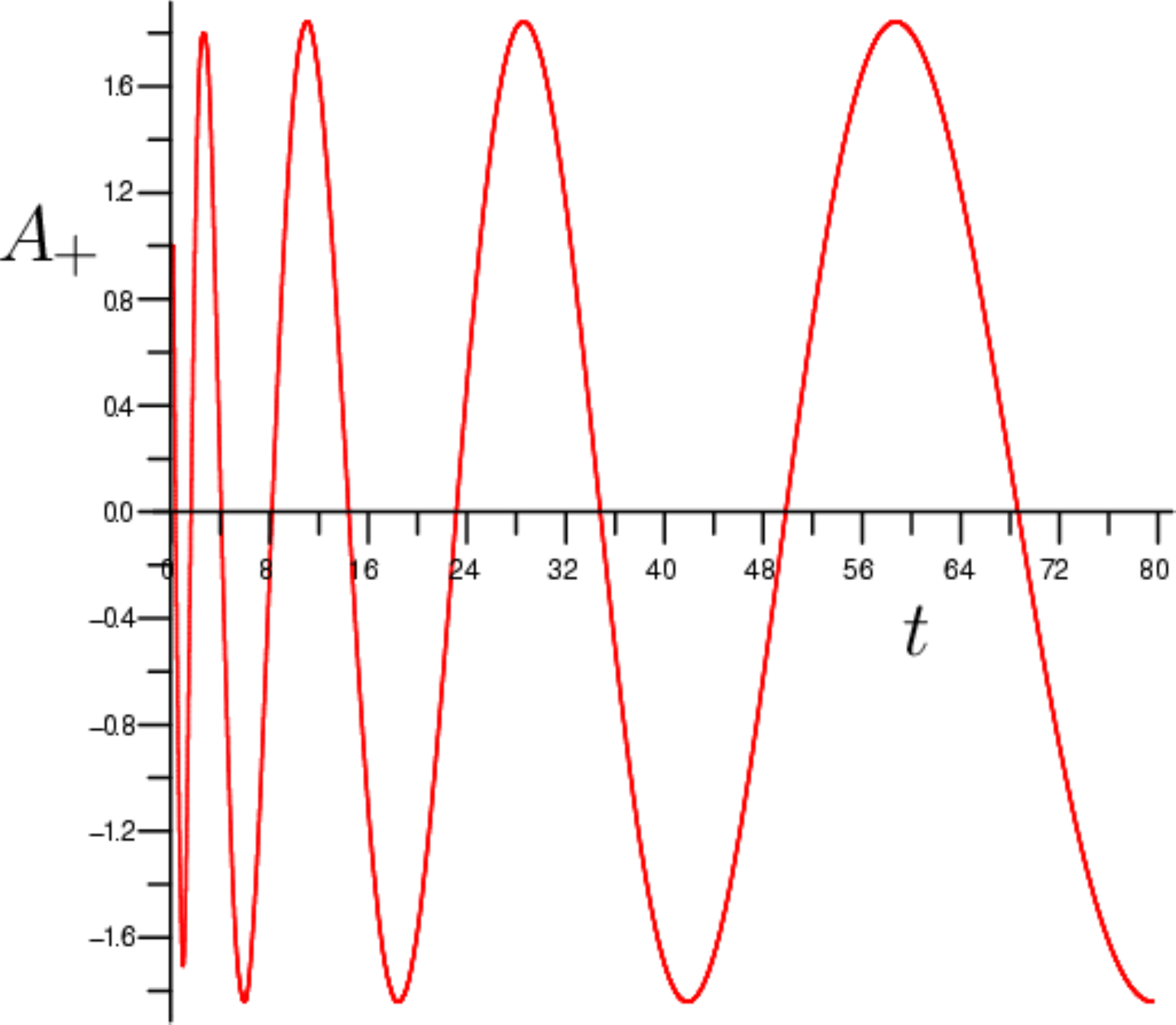}
  \caption{Numerical solution of the system of equations (\ref{eq:H}), (\ref{eq:phi and A}) and (\ref{eq:A}) (without back-reaction) for $k=m_\phi$. There is no amplification of the mode function, which is well-described by the approximate solution (\ref{eq:A+}). }
  \label{fig:A+_stable}
\end{figure}

Equation (\ref{eq:growth}) hence implies that preheating into gauge fields is very effective for $\phi_\mathrm{e}/F\gg 1$. However, this is not the regime that applies in natural models of inflation, where $\phi_\mathrm{e}/F$ is of order one (a small number for our purposes). In order to estimate the growth in the opposite regime, we shall set $\epsilon\propto \phi_\mathrm{e}/F\to 0$.  In that case,  the solution of the Mathieu equation has the growing mode
\be\label{eq:A+}
	\tilde{A}_+ \propto t^{1/3}
	\cos\left(\frac{6\, k \, t_\mathrm{e}^{2/3}}{m_\phi a_\mathrm{e}}t^{1/3} + \varphi\right),
\ee
which implies that the original variable $A_+$ oscillates with constant amplitude. Therefore,  there is no parametric resonance in the decay of the inflaton into gauge fields. As before, the lack of parametric resonance is due to the  modest value of $\phi_\mathrm{e}/F$, which is a reflection of the weak couplings between the inflaton and matter. Numerical investigation including the effects of expansion confirms this.  Fig.~(\ref{fig:A+_stable}) shows for instance the evolution of $A_+$ for $k=m_\phi$, which is very well approximated by equation (\ref{eq:A+}). As clearly seen in the figure, there is no particle production. 

Because the equation of motion is invariant under $t\to t+\pi$ and $\epsilon\to -\epsilon$, similar results hold for $A_-$ modes.  Notice that the comoving number density is now given by
\be
	n_{\bf k}^{(\pm)} 
	= \frac{a^2}{2\omega_k}
          \left(|\dot{A}_\pm|^2 + \left(\frac{k}{a}\right)^2|A_\pm|^2\right)-\frac{1}{2}.\label{eq:ndensity+}  
\ee
%==================================================================================================

\section{Conclusions}
It is obviously necessary to repopulate the universe with matter after a period of early universe inflation. The detailed process through which this takes place can be of great significance since, although a thermally equilibrated bath ultimately results, non-equilibrium relics from preheating can have important cosmological consequences, even if the ultimate reheat temperature of the universe is below the mass scale of any relics of interest.

The possible implications of preheating include the overclosure of the universe through the production of monopoles, moduli or gravitinos, and new possibilities for the generation of the baryon asymmetry of the universe, either at the Grand Unified (GUT)~\cite{Kolb:1996jt,Kolb:1998he} or electroweak~\cite{Krauss:1999ng,Garcia-Bellido:1999sv,Copeland:2001qw,Smit:2002yg}
scale. It is therefore important to examine the different ways in which parametric resonance may take place in inflationary models.

Because phenomenologically viable inflation requires extremely flat potentials, models in which the couplings of the inflaton to itself and to other fields are naturally suppressed are particularly attractive. One way to achieve this is to construct the inflaton as a pseudo-Nambu-Goldstone boson of a spontaneously broken $U(1)$ symmetry---{\it natural inflation}. The $U(1)$ symmetry is then realized as a shift symmetry on the inflaton field, and explicit soft breaking terms render this symmetry approximate, and generate a naturally small mass and approximately flat potential for the inflaton. A consequence of the residual approximate symmetry is that direct couplings o the inflaton to matter fields are correspondingly suppressed, leading to the interesting possibility that derivative couplings, unconstrained by a shift symmetry, may be the dominant type of coupling to matter.

In this paper, we have used analytic and numerical techniques to study reheating in models in which derivative couplings between the inflaton and matter fields are expected to play the dominant role---such as in natural inflation.  We have seen that successful reheating places non-trivial constraints on these models and confronts them with  several challenges.  In particular, for heavy scalar fields ($m_\chi\gg m_\phi$) parametric resonance is ineffective in producing matter particles, and the perturbative decay is forbidden on kinematic grounds. For light scalars  ($m_\chi\ll m_\phi$) there is a  a novel long-wavelength instability that causes the zero mode of matter to grow as a power law, but in this case,  the production of fluctuations of matter during inflation leads to several cosmological problems. We note that we have not addressed any mechanism to keep matter fields light, which could lead to further constraints.

Some of these problems are avoided when the inflaton couples derivatively to gauge fields, as in axion-like models.  Because of the gauge symmetry, these vector fields are automatically light, thus allowing the perturbative decay of the inflaton. At the same time, the conformal nature of their couplings to gravity circumvents all the problems associated with the presence of light scalars during inflation. Nevertheless, we have found that parametric resonance is absent for  the values of parameters implied by natural inflationary models. As for couplings to scalars, the origin of this absence is the weakness of the coupling of the inflaton to matter, which is characterized by the ratio of the inflaton at the end of inflation to  the spontaneous breaking scale, and is hence of order one at most.

%==================================================================================================

\begin{acknowledgments}

The work of CAP was supported in part by the National Science Foundation (NSF) under grant PHY-0604760. The work of MT and EW was supported in part by the NSF under grant PHY-0354990, by funds from Syracuse University and by Research Corporation.

\end{acknowledgments}

%==================================================================================================

%==================================================================================================

%==================================================================================================

\end{document}